\documentclass[10pt,twocolumn,letterpaper]{article}
\usepackage[pagenumbers]{cvpr}

\definecolor{cvprblue}{rgb}{0.21,0.49,0.74}
\usepackage[pagebackref,breaklinks,colorlinks,allcolors=cvprblue]{hyperref}

\def\paperID{21845}
\def\confName{CVPR}
\def\confYear{2026}
\usepackage{multirow}
\title{Style-Instructed Mask-Free Virtual Try On}

\author{Mengqi Zhang$^{1,2}$\thanks{Work done during internship at Amazon.} \quad Qi Li$^{1}$ \quad Mehmet Saygin Seyfioglu$^1$ \quad Karim Bouyarmane$^1$\\
$^1$Amazon \quad $^2$ Georgia Institute of Technology
}

\begin{document}
\makeatletter
\let\@oldmaketitle\@maketitle
\renewcommand{\@maketitle}{\@oldmaketitle
    \centering
    \includegraphics[width=\linewidth]{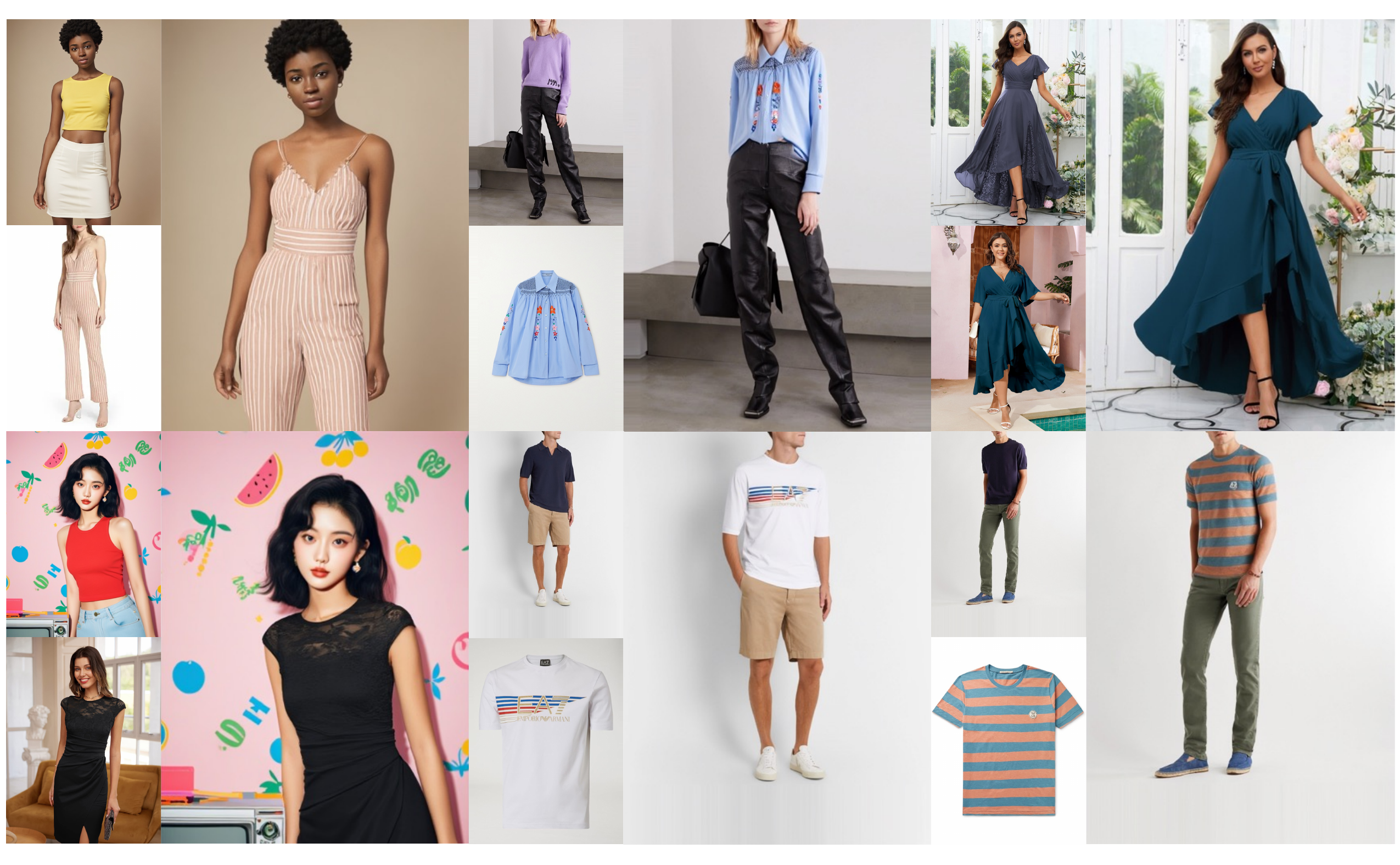}
    \captionof{figure}{\small We propose a framework that enables high-fidelity garment transfer and fine-grained stylistic control, maintaining visual consistency across diverse human identities, complex poses, and a broad spectrum of clothing categories.}
    \label{fig:teaser}
\bigskip}
\makeatother
\maketitle
\begin{abstract}
Virtual Try-On is a promising research area with broad applications in e-commerce and everyday life, enabling users to visualize garments on themselves or others before purchase. Most existing methods depend on predefined or user-specified masks to guide garment placement, but their performance is highly sensitive to mask quality, often causing misalignment or artifacts, and introduces redundant steps for users. To overcome these limitations, we propose a mask-free virtual try-on framework that requires only minimal modifications to the underlying architecture while remaining compatible with common diffusion-based pipelines. To address the increased ambiguity in the absence of masks, we integrate an attention-based guidance mechanism that explicitly directs the model to focus on the target garment region and improves correspondence between the garment and the person. Additionally, we incorporate instruction prompts, allowing users to flexibly control garment categories and wearing styles, addressing the underutilization of prompts in prior work and improving interaction flexibility. Both qualitative and quantitative evaluations across multiple datasets demonstrate that our approach consistently outperforms existing methods, producing more accurate, robust, and user-friendly try-on results. Project page: \href{https://smf-vto.github.io}{https://smf-vto.github.io}.
\end{abstract}  
\vspace{-3em}
\section{Introduction}
\label{sec:intro}

Virtual Try-On (VTO) has attracted increasing attention in the computer vision area due to its growing potential in e-commerce and daily-life applications. Given a target garment, VTO aims to realistically place and visualize it on a person, benefiting from recent advances in generative modeling, especially conditional diffusion-based generation.

Benefiting from diffusion models and conditional modeling~\citep{ho2020denoising,zhang2023adding,mou2024t2i}, most previous approaches build mask-based architectures. Under the commonly available training setup, the data typically provides paired samples of a target garment image and a ground-truth person image wearing the same garment. However, at inference time, the desired input often includes the same person wearing the original garment, which is generally unavailable during training. To handle this mismatch, many methods~\citep{chong2024catvton,li2025dit} formulate VTO as an inpainting problem: they mask out the target clothing region on the person image and use the masked image as the source input to synthesize the try-on result conditioned on the target garment. While effective, this pipeline is highly sensitive to the mask quality, particularly during the inference time, and this may introduce redundant user effort when masks need to be manually drawn.

To streamline the workflow, several recent works~\citep{morelli2023ladi,choi2021viton, li2025dit,han2025instructvton,li2025efficient,xu2025deft} propose adding an additional mask module to automatically predict masks at inference. This reduces manual labor and further facilitates the use of mask-based pipelines, which remain dominant in current VTO systems. However, extensive observations reveal several persistent drawbacks: (1) The try-on quality remains tightly coupled with mask accuracy, especially for automatically predicted masks, where errors can lead to noticeable artifacts such as leakage when the mask region is incorrectly estimated. (2) The issue becomes more pronounced in cross-category try-on. For instance, when a person wearing a short-sleeved shirt tries on a long-sleeved shirt, the mask predictor may only remove the short-sleeve region while leaving the arms exposed, which often causes the model to "paste" the long-sleeve appearance into the masked area without properly reasoning about the underlying structure, resulting in unnatural synthesis.

Motivated by these failure cases, we propose a mask-free VTO framework with text instructions to provide additional control over garment type and wearing style. Compared to existing mask-free approaches~\citep{zhang2025boow,chong2024catvton}, we supervise the network without additional control modules, which requires only minimal architectural changes. Nevertheless, removing masks increases ambiguity: without explicit spatial guidance, the model may generate incorrect garment attributes (e.g., producing a short dress when the target garment is a long dress) or yield undesirable try-on results. To mitigate this, we incorporate text instruction prompts into the pipeline to assist the model in correcting such mistakes, enabling seamless multi-garment try-on and supporting multi-turn editing. In addition, to stabilize training and encourage the model to focus on the target garment region, we introduce attention-layer guidance as an auxiliary loss, which provides an effective performance boost. 

\noindent In summary, our contributions are:
\begin{itemize}
    \item We introduce a mask-free framework for VTO with minimal architectural modifications, achieving superior performance compared to prior works and remaining competitive against strong mask-based models.
    \item To reduce ambiguity after removing masks, we integrate an auxiliary loss on attention layers that encourages the model to attend to the target garment region, leading to clear improvements in try-on quality.
    \item We incorporate text instruction prompts during training to enable additional control over garment type and wearing style, and to support seamless multi-garment try-on and interactive multi-turn editing.
\end{itemize}

\section{Related Work}
\label{sec:related_work}

Virtual try-on (VTO) has advanced rapidly in recent years, driven by its commercial value and its ability to provide personalized garment visualization. Existing methods are commonly grouped into mask-based, mask-free, and text-guided paradigms. Across these directions, diffusion models have recently become a strong backbone, improving synthesis fidelity and enabling more flexible conditional control.

\textbf{Mask-based Virtual Try-On.} Early VTO systems typically rely on human parsing or segmentation maps as spatial priors to guide garment warping and subsequent image synthesis. VITON~\citep{han2018viton} first introduced a coarse-to-fine pipeline that warps garments using segmentation cues, and CP-VTON~\citep{wang2018toward} extended this design by learning dense conditional warping flows. ClothFlow~\citep{han2019clothflow} and CP-VTON+~\citep{minar2020cp} further refine deformation and preserve garment boundaries through flow-based modeling and edge-aware objectives. While PF-AFN~\citep{ge2021parser} removes explicit parsing, it still depends on pose-based and appearance-based flows to align garments with body shape. In general, these pipelines can produce sharp results when the underlying parsing is reliable, but their performance often degrades noticeably under segmentation errors. In addition, dependence on auxiliary parsing/segmentation networks increases system complexity and can affect inference efficiency and robustness. Recent diffusion-based VTO approaches largely inherit this mask-based formulation: although diffusion improves generation capacity, segmentation inaccuracies may still propagate into the synthesis stage and impair garment-body alignment.

\textbf{Mask-free Virtual Try-On.} To reduce reliance on external annotations, mask-free methods aim to learn garment-body correspondence directly from raw images. ACGPN~\citep{yang2020towards} proposes adaptive parsing generation to guide synthesis without requiring fixed masks at inference. StylePoseGAN~\citep{sarkar2021style} leverages pretrained GAN priors and latent control to disentangle garment and pose factors without explicit structural inputs. M3D-VTON~\citep{zhao2021m3d} incorporates 3D cues such as pose and depth to improve robustness under occlusion and viewpoint changes. More recently, mask-free diffusion-based designs have shown encouraging results. TryOnDiffusion~\citep{zhu2023tryondiffusion} synthesizes try-on images by conditioning on pose and garment representations without explicit segmentation or warping, and DCTON~\citep{ge2021disentangled} improves deformation realism in a mask-free setting by modeling body-garment constraints within the diffusion process. Overall, mask-free pipelines can simplify the system and avoid error-prone parsing components, but the absence of explicit garment localization often leads to spatial ambiguity, especially for loose-fitting garments or heavily occluded regions.

\textbf{Text-Guided Virtual Try-On.} Introducing natural language provides a more intuitive interface for controlling try-on outputs. While early VTON methods are primarily image-driven, several recent works explore language as an additional conditioning signal. Text-based editing techniques developed in the GAN literature (e.g., StyleCLIP~\citep{patashnik2021styleclip} and TediGAN~\citep{xia2021tedigan}) demonstrate that prompts can guide semantic edits, though these methods are not tailored to try-on. In the diffusion regime, TryOnDiffusion~\citep{zhu2023tryondiffusion} investigates free-form descriptions as conditioning (e.g., garment type and color). Related directions such as Text2Human~\citep{jiang2022text2human} further show that combining language with structural conditions can produce high-quality human synthesis. Despite these advances, language is still underutilized in many VTO systems: prompts are often treated as a weak or coarse control input, and relatively few methods support fine-grained, editable control over both garment type and wearing style in a unified manner.

Motivated by the strengths and limitations of mask-free and text-guided paradigms, we propose a diffusion-based VTO framework that removes the need for parsing or segmentation masks while introducing only minimal architectural changes. To mitigate the spatial ambiguity inherent to mask-free designs, we incorporate attention-based garment localization during training to encourage the model to focus on the relevant garment region. We further enable controllable try-on through natural language prompts that describe both garment identity (e.g., ``denim jacket'', ``floral dress'') and style attributes (e.g., ``oversized'', ``sleeveless''). Compared to prior methods that either depend on structural priors or treat language as a secondary signal, our framework integrates mask-free generation, prompt-driven control, and attention-guided spatial reasoning within the diffusion pipeline.
\section{Method}
\label{sec:method}
\begin{figure}[t]
    \centering
    \includegraphics[width=\linewidth]{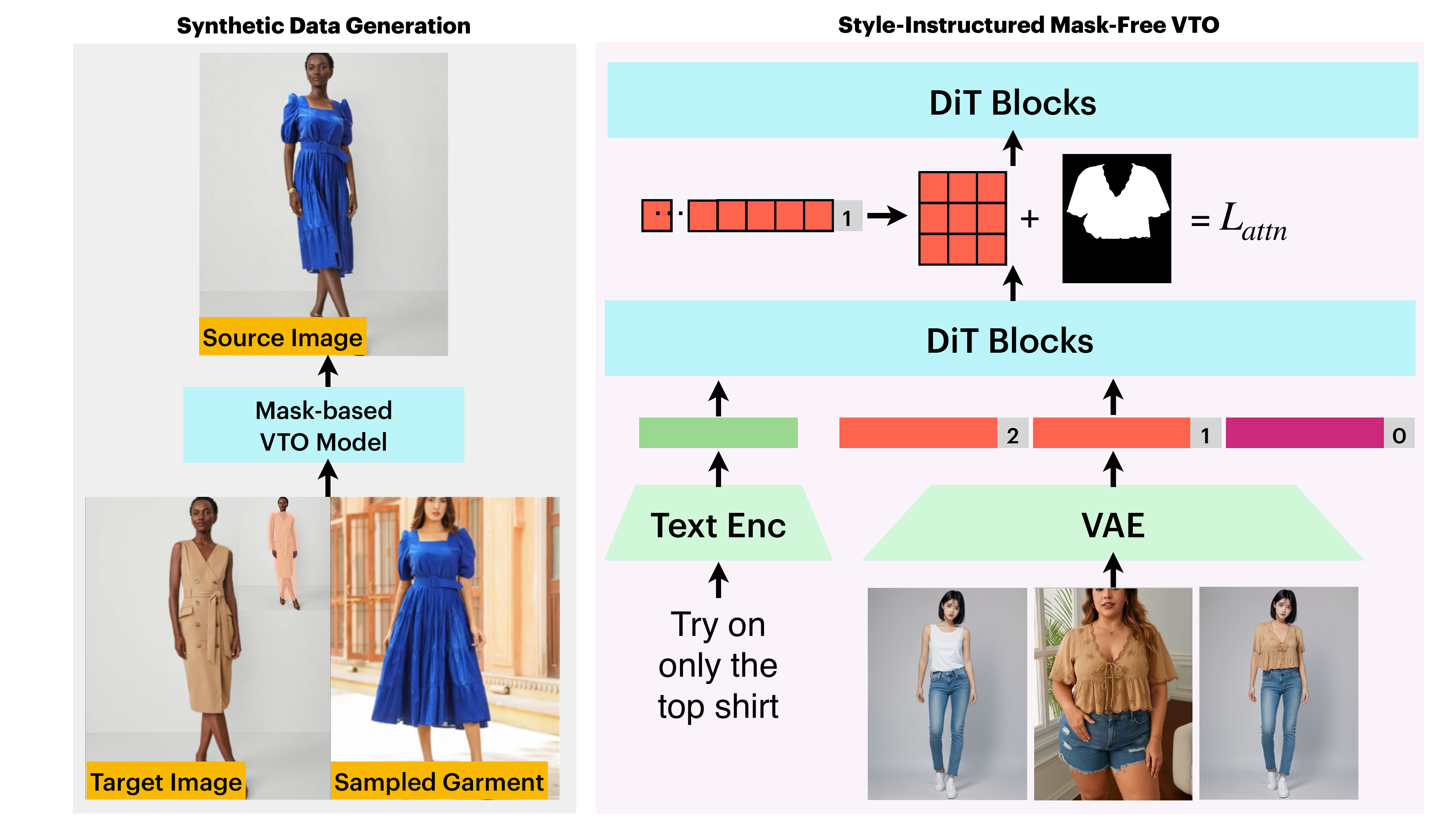}
     \caption{\textbf{Overview of the proposed Style-Instructed Mask-Free Virtual Try-On (SMF-VTO)}. Left: Synthetic data generation. A pretrained mask-based VTO model synthesizes a source image by combining a target person image with a sampled alternate outfit, producing source–target pairs that enable triplet-style supervision without manual segmentation. Right: SMF-VTO architecture. Person/garment images are encoded by a VAE, while textual style instructions are encoded by a text encoder; the resulting tokens are fused and processed by DiT blocks. We further apply an attention guidance loss, $L_{\text{attn}}$, to encourage the model’s internal attention to concentrate on garment regions, improving spatial fidelity and controllability in a fully mask-free setting.}
    \label{fig:method}
\end{figure}
SMF-VTO introduces a mask-free pipeline with additional text guidance to mitigate the problems brought by previous mask-based approaches. First, a preliminary overview is introduced to denote the symbols and define the problem space. Then, the details of the framework are elaborated to explain several essential components in our approach. Finally, we would introduce the inference time details of the work.

\subsection{Preliminary and Denotation}

In this section, we briefly review the flow-matching formulation used in our generative model, and then define the VTO problem setting and notations used throughout the paper.

\paragraph{Flow-matching diffusion.} Diffusion models are powerful generative frameworks that learn to invert a known noising process. Recently, flow-matching diffusion has emerged as a deterministic alternative that learns a time-dependent velocity field, enabling sampling via ODE integration.

Let $\mathbf{x} \in \mathbb{R}^d$ denote a sample from the data distribution $p_{\text{data}}(\mathbf{x})$. The forward process defines a trajectory $\mathbf{x}_t$ for $t\in[0,1]$ via linear interpolation between clean data and noise:
\vspace{-1mm}
\begin{equation}
\mathbf{x}_t = (1-t)\mathbf{x}_0 + t\mathbf{z}, \quad \mathbf{z} \sim p_{\text{prior}}(\mathbf{z}),
\end{equation}
\vspace{-1mm}
where $\mathbf{z}$ is typically drawn from a standard Gaussian prior.
Flow matching learns a velocity field $v_\theta(\mathbf{x}_t, t)$ such that solving
\vspace{-0.5mm}
\begin{equation}
\frac{d\mathbf{x}_t}{dt} = v_\theta(\mathbf{x}_t, t)
\end{equation}
\vspace{-0.5mm}transports noise $\mathbf{z}$ to data $\mathbf{x}_0$.

Instead of learning a stochastic score function as in DDPMs, flow matching minimizes a closed-form objective that directly supervises the velocity field. The target flow is
\vspace{-0.5mm}
\begin{equation}
v^*(\mathbf{x}_t,t) = \frac{d\mathbf{x}_t}{dt} = \mathbb{E}[\mathbf{x}_0-\mathbf{z}\mid \mathbf{x}_t,t],
\end{equation}
\vspace{-0.5mm}which is approximated using the known interpolation, yielding the standard training loss:
\vspace{-0.5mm}
\begin{equation}
\mathcal{L}_{\text{FM}}=
\mathbb{E}_{\mathbf{x}_0,\mathbf{z},t}\left[
\left\|v_\theta(\mathbf{x}_t,t)-(\mathbf{x}_0-\mathbf{z})\right\|_2^2
\right],
\end{equation}
\vspace{-0.5mm}where $t\sim\mathcal{U}(0,1)$.
This formulation supports efficient deterministic sampling via ODE integration and is well-suited for conditional generation in VTO.

\paragraph{Problem statement.}
Given a person image $\mathbf{I}_p\in\mathbb{R}^{H\times W\times 3}$ and a target garment image $\mathbf{I}_g\in\mathbb{R}^{H\times W\times 3}$, the goal is to synthesize a try-on image $\mathbf{I}_t\in\mathbb{R}^{H\times W\times 3}$ where the person in $\mathbf{I}_p$ realistically wears the garment from $\mathbf{I}_g$, while preserving pose, identity, and background. We optionally condition on auxiliary signals such as pose $\mathbf{P}$ and a text instruction $\mathbf{T}$ (e.g., ``a red long-sleeved dress'') to improve semantic control. We define a conditional generator $\mathcal{G}_\theta$ as
\begin{equation}
\mathbf{I}_t=\mathcal{G}_\theta(\mathbf{I}_p,\mathbf{I}_g,\mathbf{P},\mathbf{T}),
\end{equation}
where $\theta$ denotes learnable parameters. The synthesized image $\mathbf{I}_t$ should satisfy: (i) garment fidelity (preserving appearance from $\mathbf{I}_g$ and / or semantics from $\mathbf{T}$), (ii) identity and pose preservation, (iii) visual realism, and (iv) controllability through text instructions.

Unlike traditional mask-based methods that use segmentation masks $\mathbf{M}_p$ of the person to guide localization and warping, SMF-VTO targets a mask-free setting at inference time: the model must learn to align and render garments without requiring user-provided or externally predicted masks.

\subsection{Mask-Free Virtual Try-On in SMF-VTO}
\label{sec:maskfree}

In contrast to prior VTO pipelines that depend on explicit segmentation masks or human parsing maps to guide garment placement~\citep{han2018viton,wang2018toward}, SMF-VTO adopts a fully \emph{mask-free} formulation. The model receives no pixel-level masks for either the person or the garment at inference. Instead, it learns garment localization and alignment implicitly from image-level supervision and conditional inputs.

This setting introduces two key challenges: (1) without explicit spatial priors, garment--body correspondence becomes ambiguous; and (2) without semantic boundaries, preserving fine-grained garment structure (e.g., sleeves, collars, and layered regions) becomes more difficult.

SMF-VTO addresses these issues through two complementary mechanisms. First, we rely on strong conditional modeling: the model is conditioned on the image pair $(\mathbf{I}_p,\mathbf{I}_g)$, the pose embedding $\mathbf{P}$, and a text instruction $\mathbf{T}$ describing garment category and style (described in Sec.~\ref{sec:text}). Second, we introduce an explicit attention-guided objective (Sec.~\ref{sec:attn}) that encourages the diffusion backbone to allocate capacity to garment-relevant regions, improving spatial precision without requiring masks at inference.

\subsection{Style Control via Text Instructions}
\label{sec:text}

To enable flexible and fine-grained control, SMF-VTO incorporates natural language instructions as a high-level guidance signal. Users can specify garment attributes (e.g., ``long-sleeved blouse'', ``floral dress'', ``oversized hoodie'') and wearing styles (e.g., ``tucked in'', ``off-shoulder'', ``open jacket'') through a free-form prompt $\mathbf{T}$.

We encode $\mathbf{T}$ using a pretrained text encoder such as CLIP~\cite{radford2021learning} or T5~\citep{raffel2020exploring}, producing a text embedding $\mathbf{e}_T\in\mathbb{R}^d$. The embedding is projected with a learnable MLP and injected into the generative model via cross-attention, serving as a global semantic condition. In addition, we concatenate the projected text tokens with pose and appearance tokens in the conditioning module, providing complementary semantic priors that improve alignment between the instruction and the synthesized output.

By leveraging compositional text prompts, SMF-VTO supports controllable try-on across a broad range of garment categories and styles, including combinations not explicitly seen during training. This design enables interactive editing and improves usability without requiring any explicit masks.

\begin{table*}[!t]
\caption{\textbf{Quantitative performance of DressCode.}
We use DressCode as the test set. We compare mask-based methods (LaDI-VTON, OOTD, IDM-VTON) with our mask-free \textbf{SMF-VTO}.}
\label{tab:dresscode-unpaired}
\centering
\scalebox{0.85}{
\begin{tabular}{lcccc}
\toprule
\textbf{DressCode (Unpaired)} & SSIM $\uparrow$ & LPIPS $\downarrow$ & FID $\downarrow$ & IS $\uparrow$\\
\midrule
LaDI-VTON~\cite{morelli2023ladi} & 0.7656 & 0.2366 & 11.08 & - \\
OOTD~\cite{xu2025ootdiffusion} & 0.8854 & 0.0533 & 12.567 & - \\
IDM-VTON~\cite{choi2024improving} & 0.8797 & 0.0563 & 9.546 & - \\
\midrule
\textbf{SMF-VTO} & 0.9318 & 0.0731 & 3.7960 & 3.9445 \\
\bottomrule
\end{tabular}}
\end{table*}

\begin{table*}[!t]
\caption{\textbf{Quantitative performance of VITON-HD.}
We use VITON-HD as a test set. \textbf{SMF-VTO} achieves strong overall performance across most metrics.}
\label{tab:vitonhd-unpaired}
\centering
\scalebox{0.85}{
\begin{tabular}{lcccc}
\toprule
\textbf{VITON-HD (Unpaired)} & SSIM $\uparrow$ & LPIPS $\downarrow$ & FID $\downarrow$ & IS $\uparrow$\\
\midrule
DCI-VTON~\cite{gou2023taming} & 0.8545 & 0.1800 & 8.998 & - \\
LaDI-VTON~\cite{morelli2023ladi} & 0.8395 & 0.2014 & 11.08 & - \\
CAT-DM~\cite{zeng2024cat} & 0.8391 & 0.1621 & 10.28 & - \\
StableVITON~\cite{kim2024stableviton} & 0.8519 & 0.1479 & 9.851 & - \\
OOTD~\cite{xu2025ootdiffusion} & 0.8301 & 0.1420 & 12.19 & - \\
IDM-VTON~\cite{choi2024improving} & 0.8547 & 0.1223 & 9.265 & - \\
\midrule
BooW-VTON~\cite{zhang2025boow} & 0.8618 & 0.1080 & 8.809 & - \\
\textbf{SMF-VTO} & 0.8904 & 0.1265 & 7.916 & 3.0659 \\
\bottomrule
\end{tabular}}
\end{table*}

\subsection{Attention-Guided Auxiliary Loss}
\label{sec:attn}

To improve spatial focus in the absence of explicit masks, we introduce an attention-guided auxiliary objective that regularizes internal attention maps in the diffusion backbone. Rather than supervising attention with external dense annotations, we encourage the model's attention to concentrate on garment-relevant regions using a weak localization prior available during training.

Let $A_l\in\mathbb{R}^{H_p\times W_p}$ denote an attention map extracted from the $l$-th DiT (transformer) block in patch space. We reproject $A_l$ to align with the latent token grid and define an auxiliary loss that increases attention responses within garment-relevant regions:
\begin{equation}
\mathcal{L}_{\text{attn}}
= -\frac{1}{L}\sum_{l=1}^{L} \left\| A_l \odot M \right\|_1,
\end{equation}
where $M$ is a training-time garment region indicator (e.g., obtained from weak localization or pseudo labels) and $\odot$ denotes element-wise multiplication. Intuitively, this objective encourages the model to allocate higher attention mass to garment regions across layers, acting as a soft focus regularizer and improving garment localization.

The final training objective combines the flow-matching loss with the attention objective:
\begin{equation}
\mathcal{L}_{\text{total}}
=
\mathcal{L}_{\text{FM}}
+
\lambda_{\text{attn}}\,\mathcal{L}_{\text{attn}},
\end{equation}
where $\lambda_{\text{attn}}$ controls the strength of attention regularization. Importantly, $M$ is only used during training; SMF-VTO remains mask-free at inference.

\section{Experiment}

\label{sec:exp}

\subsection{Data Generation Pipeline}
\label{sec:data}

Training a mask-free try-on model requires supervision that matches the inference setting, i.e., a source person image and a separate target garment. However, standard VTO datasets typically provide paired samples where the person in the target image already wears the corresponding in-shop garment. To bridge this gap, we construct synthetic training triplets $(\mathbf{I}_p, \mathbf{I}_g, \mathbf{I}_t)$ for our large-scale dataset. Each triplet includes: (i) a source person image $\mathbf{I}_p$ depicting the same identity and pose as the target, but wearing a different outfit; (ii) a target garment image $\mathbf{I}_g$ (e.g., catalog / in-shop); and (iii) a ground-truth try-on target $\mathbf{I}_t$ where the person wears $\mathbf{I}_g$.

Concretely, for each paired sample $(\mathbf{I}_t, \mathbf{I}_g)$ in the dataset, we synthesize $\mathbf{I}_p$ by changing the clothing in $\mathbf{I}_t$ while preserving identity, pose, and background. We employ a pretrained CatVTON~\cite{chong2024catvton} as a teacher generator. Given $\mathbf{I}_t$ and a randomly sampled garment $\tilde{\mathbf{I}}_g \neq \mathbf{I}_g$, CatVTON produces a source image $\mathbf{I}_p$ in which the person wears $\tilde{\mathbf{I}}_g$. This procedure yields triplets where $\mathbf{I}_p$ and $\mathbf{I}_t$ share the same person and pose but differ in clothing, closely matching the deployment scenario.

To improve the quality of supervision, we filter out noisy triplets based on simple heuristics, including large pose discrepancies, severe artifacts (e.g., broken limbs, distorted garment boundaries), and obvious identity drift by using VLM to judge. The resulting triplets allow SMF-VTO to learn garment replacement conditioned on $(\mathbf{I}_p, \mathbf{I}_g)$ and, when available, a text instruction $\mathbf{T}$, without requiring any human-annotated segmentation masks. We additionally associate each triplet with a text instruction prompt $\mathbf{T}$ to enable style-conditioned training. Specifically, we also use the VLM to caption the outfit appearance in the target garment image $\mathbf{I}_t$, producing a natural-language description of garment category and style attributes (e.g., sleeve length, fit, and notable patterns). We then use this caption as the prompt $\mathbf{T}$ during training, which provides lightweight semantic supervision without any manual annotation.

\begin{figure*}[t]
    \centering
    \includegraphics[width=\linewidth]{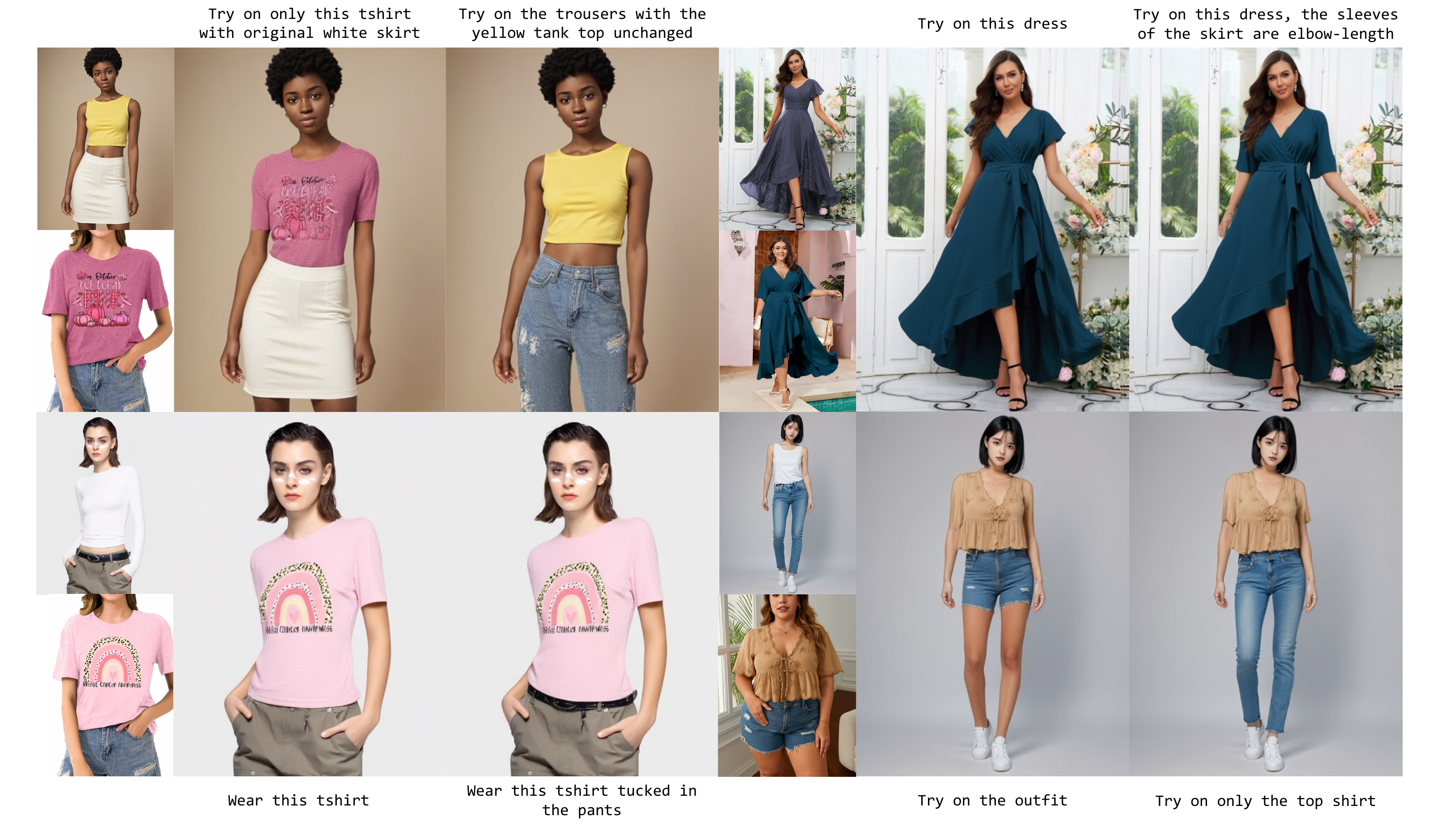}
     \caption{\textbf{Text-instructed style control results of SMF-VTO.}
Given the same source person image and pose, SMF-VTO generates different try-on results guided by natural language prompts describing garment type and wearing style (e.g., ``wear this t-shirt tucked in the pants'', ``try on this dress with elbow-length sleeves''). The results illustrate that SMF-VTO can follow fine-grained instructions while maintaining realistic garment transfer and spatial alignment, without any segmentation masks.}
    \label{fig:text}
\end{figure*}

\subsection{Implementation}
We implement our method based on the open-source \texttt{Kontext}~\citep{labs2025flux1kontextflowmatching} diffusion framework. All experiments are conducted on 8 NVIDIA H100 GPUs (80GB each), using mixed-precision training (FP16) for efficiency. We train the model for approximately 5 days, corresponding to roughly 500k steps depending on the dataset. We use the Adam optimizer with a learning rate of $1 \times 10^{-5}$, and all other training hyperparameters (including beta schedule, noise sampler, and architecture settings) follow the default configuration provided by \texttt{Kontext}. During inference, we use DDIM sampling with 25 steps unless otherwise specified.

For evaluation, we report results under the \emph{unpaired} setting, where person--garment pairs are randomly mismatched to assess robustness and generalization beyond the paired training distribution. We use SSIM and LPIPS to measure perceptual similarity and structure consistency, and FID/IS to quantify realism and distribution-level quality.

\subsection{Evaluation Protocol}
\label{sec:eval}

We evaluate all methods under the unpaired setting, where the person image and the target in-shop garment are intentionally mismatched. This protocol stresses generalization and robustness, and avoids overestimating performance due to dataset-specific paired correlations. Concretely, for each person image $\mathbf{I}_p$, we randomly sample a target garment $\mathbf{I}_g$ from the test split with the constraint that it is not the garment originally worn by the person.

We report both perceptual similarity and distribution-level realism metrics. SSIM and LPIPS measure structural consistency and perceptual distance between the synthesized image and the ground-truth target image when available, while FID quantifies the realism of generated samples by comparing their feature distributions to real images. We additionally report IS when the baseline provides it. Since unpaired evaluation can be sensitive to pairing strategy (e.g., cross-category mismatch), we keep the same pairing list across methods to ensure a controlled comparison.




\subsection{Quantitative Comparison}

We present quantitative comparisons under the \emph{unpaired} evaluation setting on DressCode and VITON-HD in Tab.~\ref{tab:dresscode-unpaired} and Tab.~\ref{tab:vitonhd-unpaired} as test sets. Unpaired evaluation intentionally mismatches person--garment pairs to test robustness and generalization, and is particularly relevant for real-world try-on where the target garment may differ substantially from the original outfit.

Across both datasets, SMF-VTO achieves consistently strong performance on SSIM/LPIPS and substantially improved FID, indicating better perceptual quality and realism under challenging unpaired conditions. Compared to mask-based pipelines, which can suffer from segmentation-induced misalignment, and mask-free baselines that may exhibit spatial ambiguity, our attention-guided generation produces more coherent garment placement and more faithful textures. In addition, text-guided style control further improves flexibility, enabling diverse garment types and style attributes compared to prompt-agnostic baselines.

\subsection{Style Guidance Visualization}
\label{sec:stylevis}

We visualize the effectiveness of text-based style control in Fig.~\ref{fig:text}. Given the same source person image and pose, we vary the textual instruction $\mathbf{T}$ to specify different garment categories and style attributes, such as ``a red sleeveless dress'', ``an oversized denim jacket'', ``a short-sleeved floral blouse'', or ``a tucked-in white shirt''. SMF-VTO captures both high-level semantics (e.g., dress vs.\ jacket) and fine-grained cues (e.g., sleeve length, fit), producing coherent and visually distinct outputs.

We further test compositional prompts that combine multiple attributes (e.g., ``a black long-sleeved turtleneck under a beige blazer''), illustrating the model's ability to interpret layered descriptions and generate consistent multi-garment effects. These results highlight the flexibility of SMF-VTO in supporting language-driven customization, which is difficult to achieve with prompt-agnostic or image-only try-on pipelines.
\begin{figure}[t]
    \centering
    \includegraphics[width=\linewidth]{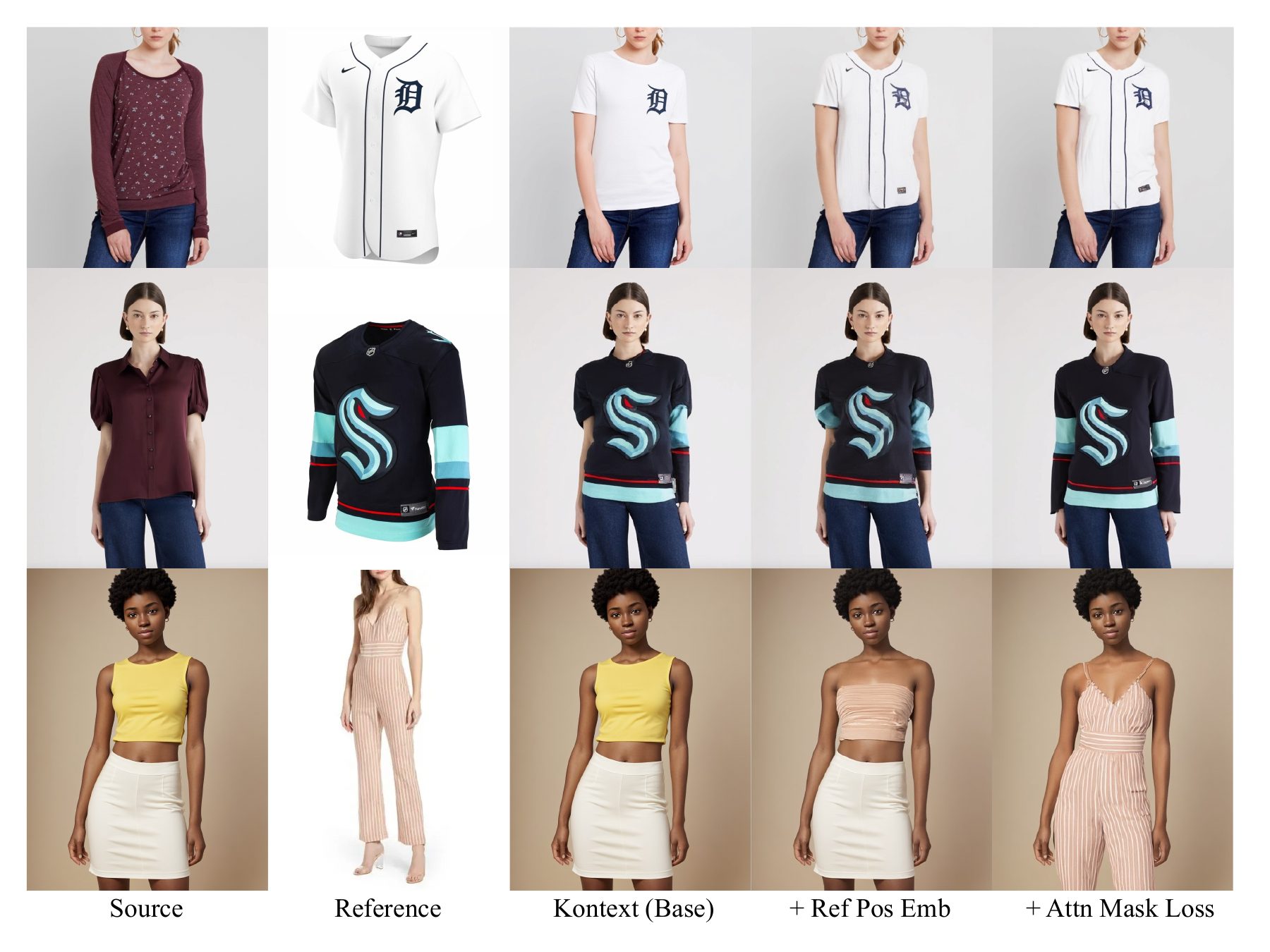}
     \caption{\textbf{Ablation study on the key components}From left to right: source person image, reference garment, baseline Kontext model, Kontext with reference positional embedding (+ Ref Pos Emb), and full model with attention-guided mask loss (+ Attn Mask Loss).The baseline struggles with incomplete garment transfer and spatial misalignment. Adding reference positional embedding improves placement and structure, while incorporating the attention-guided auxiliary loss further enhances garment detail, contour fidelity, and texture realism—demonstrating the complementary effects of both modules.}
     \vspace{-2em}
    \label{fig:abl}
\end{figure}
\section{Ablation Study}
\label{sec:abl}
\subsection{Ablation Study}

To assess the contribution of individual components in our framework, we conduct an ablation study using the \texttt{Kontext} base model as our starting point. As illustrated in Fig.~\ref{fig:abl}, we incrementally add two modules: (1) the reference positional embedding, and (2) the attention-guided mask loss. This setup allows us to analyze their impact on garment alignment, fidelity, and identity preservation.

\paragraph{Baseline (\texttt{Kontext}):}  
The base model synthesizes garments conditioned on the source person and reference garment, but struggles with coarse alignment and incomplete transfer. In the top row of Fig.~\ref{fig:abl}, the generated logo is misaligned and truncated; in the bottom row, the target garment is not faithfully transferred, and the model largely reproduces the original top.

\paragraph{+ Reference Positional Embedding:}  
We introduce a reference garment positional embedding to encode spatial priors from the target garment image. This helps the model better infer where key visual elements should appear on the body. As shown in the middle column, this module improves overall garment placement and enables better coverage of complex clothing. However, some structure ambiguity remains, particularly for fine details like straps and seams.

\paragraph{+ Attention-Guided Mask Loss:}  
Finally, we add the attention-guided auxiliary loss, which forces the model to attend to garment-relevant regions during training (see Sec.~\ref{sec:attn}). This leads to significantly improved visual fidelity and garment completeness. In the bottom row, our model correctly reproduces the striped jumpsuit with thin straps and realistic geometry, which is missing from previous versions. The attention constraint not only improves garment reconstruction, but also preserves human identity and body boundaries more effectively.

Overall, this step-wise ablation confirms that each proposed component contributes meaningfully to the final model’s spatial alignment and visual realism.

\section{Conclusion}
\label{sec:conclusion}

We propose a mask-free, diffusion-based virtual try-on framework that enables realistic and controllable garment synthesis without relying on segmentation masks. To address the spatial ambiguity of mask-free generation, we introduce an attention-guided loss and a reference positional embedding module, which significantly improve garment alignment and detail preservation. Additionally, we incorporate text-based style prompts for fine-grained control over garment type and appearance. Experiments on VITON-HD and DressCode demonstrate the effectiveness of our approach, both qualitatively and quantitatively. Future work includes extending to multi-garment composition and video try-on.
{
    \small
    \bibliographystyle{ieeenat_fullname}
    \bibliography{main}
}

\end{document}